\documentclass[letterpaper,pra,aps,floatfix,superscriptaddress,longbibliography,twocolumn,notitlepage,
accepted=2021-01-12
]{quantumarticle}
\pdfoutput=1
\usepackage[utf8]{inputenc}
\usepackage[english]{babel}
\usepackage[T1]{fontenc}
\usepackage{amsmath,amsthm,amsfonts,amssymb,xcolor}
\usepackage{graphicx}
\usepackage{microtype}
\usepackage{braket}
\usepackage[numbers,sort&compress,square]{natbib}

\makeatletter
\def\l@subsubsection#1#2{}
\makeatother

\usepackage[colorlinks=true,citecolor=blue,linkcolor=blue]{hyperref}
\usepackage[capitalize,nameinlink]{cleveref}

\theoremstyle{definition}

\begin{document}

\title{Measurement sequences for magic state distillation}

\author{Jeongwan Haah}
\affiliation{Microsoft Quantum, Redmond, Washington, USA}
\author{Matthew B. Hastings}
\affiliation{Microsoft Quantum, Santa Barbara, California, USA}
\affiliation{Microsoft Quantum, Redmond, Washington, USA}

\begin{abstract}
Magic state distillation uses special codes to suppress errors in input states,
which are often tailored to a Clifford-twirled error model.
We present detailed measurement sequences for magic state distillation protocols
which can suppress arbitrary errors on any part of a protocol,
assuming the independence of errors across qubits.
Provided with input magic states,
our protocol operates on a two-dimensional square grid
by measurements of $ZZ$ on horizontal pairs of qubits,
$XX$ on vertical pairs, and $Z,X$ on single qubits.
\end{abstract}

\maketitle

\section{Introduction}

Magic states are ancilla states that enable universal quantum computation
using only Clifford gates.
Important examples are the one-qubit state $\ket T = \ket 0 + e^{i\pi /4} \ket 1$
for general $SU(2)$ rotation
and the three-qubit state $\ket{CCZ} = \sum_{a,b,c=0,1} (-1)^{abc} \ket{a,b,c}$
for quantum coherent arithmetic.
Here we present specific measurement sequences
to distill higher fidelity $T$ and $CCZ$ states from noisy $T$ states.

We present three protocols,
each of which is a tailored implementation of a known abstract distillation scheme.
Two $T$ distillation protocols below are based on the idea of measuring the Clifford (not Pauli) stabilizer
of $T$ states~\cite{Knill2004a,Knill2004b,AliferisGottesmanPreskill2005,HHPW}.
A $CCZ$ distillation protocol below uses a generalization~\cite{campbell2017unified} of triorthogonal codes~\cite{bh},
which happens to be closely related to protocols in \cite{eastin2013distilling,jones2013low}.

Earlier considerations assumed perfect Clifford operations
with Clifford twirled noise model on nonClifford operations
and used a certain CSS code only to detect $Z$ errors.
It was known that these assumptions can be relaxed~\cite{AliferisGottesmanPreskill2005,JYHL2012,Brooks2013},
but only recently~\cite{Litinski2019,Chamberland2018,Chamberland2020}
it has become more serious to use the full or partial potential of error correction by the outermost%
\footnote{
By an \emph{outer} code we mean a code whose constituent qubits are logical qubits of some inner code.
We often consider the inner code to be a surface code.
}
code,
which is a normal weakly selfdual CSS code~\cite{HHPW} or a triorthogonal code~\cite{bh}.

Chamberland et al.~\cite{Chamberland2018,Chamberland2020}
use small instances of color codes for 
both usual error correction and Clifford stabilizer measurements.
These implementations require a degree of connectivity
that may be quite nontrivial for a two-dimensional grid of qubits in order
to accommodate ``flag'' ancillas for fault-tolerant syndrome measurements.
In addition, their implementation gives a distilled magic state encoded in a patch of color code.
We would prefer a surface code to a color code for the surface code's smaller weight of stabilizers.

Litinski~\cite{Litinski2019} focuses on a surface code architecture
and shrinks surface code patch size wherever possible to reduce the overhead.
This scheme uses the outer code to detect $Z$ errors only.
Nonrectangular, nonconvex patches are extensively used to connect rectangular (nonsquare) patches,
but the error rate estimation for those nonrectangular patches 
is a heuristic extrapolation of Monte Carlo simulation results on square patches.
If we use the outermost code of distillation as a full quantum error correcting code,
it is very important to understand the constituent qubits' error channels;
with outermost codes of distance~$3$ or less, as in~\cite{Litinski2019},
any correlated errors would invalidate the error analysis based on independent noise models.
In particular, the nature of logical error channel on surface code patches of irregular shapes 
that are used to connect rectangular logical patches,
deserves further study in regards to the protocols in \cite{Litinski2019}.
Thus, we are motivated to design protocols such that
they only require limited connectivity, limited low level operations,
and minimal assumption on error models.

The theme to use an outermost code in a distillation protocol 
as a full quantum error correcting code
continues in our design.
In contrast to \cite{Litinski2019}, we use the ability of the outer error correcting code 
to correct both $X$ and $Z$ errors with attention to correlated errors;
in contrast to \cite{Chamberland2018,Chamberland2020}, 
we use a concatenation of an inner and outer code instead of a single color code,
potentially enabling us to exploit the improved error correcting properties of the surface code or other inner code,
with consideration of limited connectivity.
Our protocols produce a distilled magic state on a standalone ``output'' qubit (or three for the $CCZ$ state)
that may be a surface code patch.
We adhere to a scenario where there is strict limitation on the elementary gates.
Namely, the elementary operations are horizontal $ZZ$ and vertical $XX$ measurements
across nearest neighbor qubits on a square grid of qubits,
along with single-qubit $X$ and $Z$ measurements.
An exception is given to output qubits which may be a bigger surface code patch.
No Hadamard gate will be used.
We imagine that poor quality magic states are handed over from a previous round of distillation.
This would be a minimal requirement in a surface code architecture,
and certainly possible using lattice surgery~\cite{surgery}.
Our protocols apply straightforwardly 
to an architecture with Majorana wires~\cite{karzig2017scalable}.

\section{Setting}

As usual, 
$X = \ket 1 \bra 0 + \ket 0 \bra 1$, 
$Z = \ket 0 \bra 0 - \ket 1 \bra 1$,
$T = \ket 0 \bra 0 + e^{i\pi/4} \ket 1 \bra 1$,
$S = T^2 = \ket 0 \bra 0 + i \ket 1 \bra 1$,
and $\ket \pm = \ket 0 \pm \ket 1$.
We drop unimportant normalizations.

\subsection{Elementary operations}

\begin{figure}[t]
\caption{
{\bf (a)} Arbitrary state teleportation by measurements,
which is possible using either a $ZZ$ or $XX$ measurement along with single-qubit measurements.
{\bf (b)} Injection of $T$ and $S$ gates by measurements.
In (a) and (b), the double line arrow means that the target Pauli correction is applied  only upon the $-1$ outcome.
{\bf (c)} Injection of $T$ gate with geometric constraints 
that only horizontal $ZZ$ and vertical $XX$ measurements are allowed
on a square grid of qubits.
The boxes with $S$ and $T$ are logical qubits containing $S$ and $T$ states, respectively.
The rest of the figure specifies what needs to be measured in order.
Since a Pauli operator is always applied by a frame update,
the drawn measurements are all that is needed on a quantum device.
The first two measurements here are those on the first figure of~(b).
If the outcome from the $ZZ$ measurement is $-1$,
then we have to apply $S$ gate by the measurements in the dashed box.
The $XX$ measurement and the two $Z$ measurement within the dashed box
are to teleport the $S$ state to the left of the data qubit according to the prescription in~(a),
and the $ZZ$ and the final $X$ measurements realize the second box of~(b).
}
\includegraphics[width=0.48\textwidth, trim = {0ex 24ex 115ex 8ex}, clip]{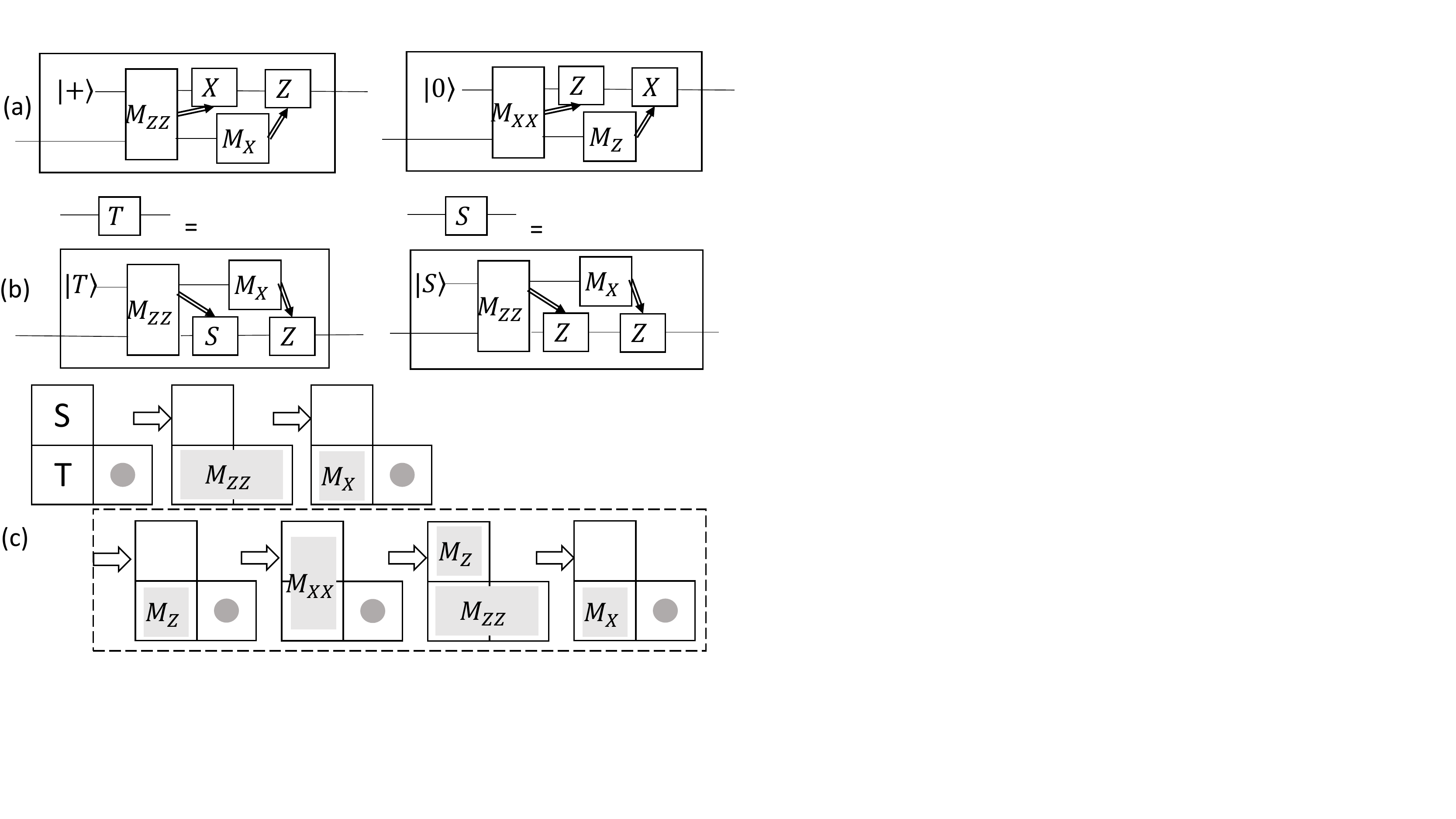}
\label{fig:injection}
\end{figure}

Every operation in the protocol that we will describe is 
a measurement of a multiqubit Pauli operator $X^{\otimes m}$ for some $m$,
a single qubit measurement in the $X$ and $Z$ basis,
and a single qubit unitary by $X,Z$ and $T$.
An exception is in the treatment of ``output'' qubits,
for which we need an embedding (a logical identity operation) 
of a qubit into a better quality qubit.
If each qubit in our protocol is a surface code patch,
this embedding amounts to growth of a patch in size.
This happens for one qubit in the entire protocol.
The single qubit unitary by $X$ and $Z$ should always be done by a Pauli frame update.

We can decompose our operations
down to single-qubit $X,Z$ measurements
and horizontal $ZZ$ and vertical $XX$ measurements across nearest neighbors.
The $T$ gate is injected via $ZZ$ and $X$ measurements.
Since we want $ZZ$ measurements to be possible only on horizontal pairs,
$T$ states need to be transported to the left of each data qubit. 
This is done by teleportation.
In \cref{fig:injection}, we specify all measurements to apply a $T$ gate.

Multiqubit $X$ measurements are performed by cat states.
Ref.~\cite{nicolas} gives a procedure
that uses $2n-1$ qubits and single-qubit $X$ and nearest neighbor $ZZ$ measurements
to produce an $n$-qubit cat state.
Using a cat state $\ket{0^{\otimes m}} + \ket{1 ^{\otimes m}}$, we can measure $X^{\otimes m}$
by taking the parity of vertical $XX$ measurements between a cat state qubit and a data qubit.
These measurements should be followed by single-qubit $Z$ measurements
on the cat state qubits and subsequent Pauli corrections on the data qubits,
to return the data qubits in the correct postmeasurement state.

Let us spell out the evolution of states during the measurement of $X^{\otimes m}$.
For any binary vector $\vec a, \vec b$ of appropriate dimension,
let $X_{\vec a}, Z_{\vec b}$ be the tensor products of $X$ and $Z$, respectively,
that have nontrivial tensor factors only on the support of $\vec a, \vec b$.
Let $\vec x$ be the binary vector of outcomes of the vertical $XX$ measurements
and $\vec z$ be the binary vector of outcomes of the $Z$ measurements on the cat state qubits.
Let $\sum_{\vec f} a_{\vec f} Z_{\vec f} \ket{+^{\otimes m}}$ be an arbitrary state on $m$ qubits.
If $m=2$, this is just an expansion of a state in the basis $\{ \ket{++}, \ket{+-}, \ket{-+},\ket{--} \}$.
\begin{align}
&\left(\sum_{\vec v:\text{even weight}} Z_{\vec v} \ket{+^{\otimes m}}\right)\otimes 
\left(\sum_{\vec f} a_{\vec f} Z_{\vec f} \ket{+^{\otimes m}}\right)\nonumber\\
\xrightarrow{M_{XX}^{\otimes m}}& 
\sum_{\vec v:\text{even}} a_{\vec v + \vec x} Z_{\vec v}\ket{+^{\otimes m}}Z_{\vec v + \vec x}\ket{+^{\otimes m}}\nonumber\\
\xrightarrow{M_{Z,\text{cat}}^{\otimes m}} &
\sum_{\vec v:\text{even}} a_{\vec v + \vec x} (-1)^{\vec v \cdot \vec z} Z_{\vec v + \vec x}\ket{+^{\otimes m}} \label{eq:XmByCat}\\
\xrightarrow{X_{\vec z,\text{data}}} &
\sum_{\vec v:\text{even}} a_{\vec v + \vec x} (-1)^{\vec x \cdot \vec z} Z_{\vec v + \vec x}\ket{+^{\otimes m}} \nonumber
\end{align}
where 
in the first line the left tensor factor is the cat state,
and in the last line $X_{\vec z}$ is the Pauli correction
that only leaves a global phase $(-1)^{\vec x \cdot \vec z}$.
Therefore, the last line equals up to a global phase
\begin{align}
&\sum_{\vec f:~|\vec f| = |\vec x| \bmod 2} a_{\vec f} Z_{\vec f}\ket{+^{\otimes m}} \\
&= \frac{1 + (-1)^{|\vec x|} X^{\otimes m}}{2} \left(\sum_{\vec f} a_{\vec f} Z_{\vec f} \ket{+^{\otimes m}}\right) \nonumber
\end{align}

\subsection{Layout}

\begin{figure}[h]
\caption{
Layout of qubits. 
The first and second rows of the box are for $T$ and $S$ states to be consumed.
The third is for an $[[n,k,d]]$ stabilizer code where circled qubits are the data qubits of the code.
The blank qubits on the third row 
are going to be used to inject $T$ by horizontal $ZZ$ measurements.
The fourth row is used to prepare cat states.
On the left of these rows, there are $k+2$ qubits of better quality,
which needs to be larger if a surface code is used to encode every qubit in this figure.
Since these better quality qubits (output) interact with the data qubits only after all $T$ states are consumed,
we can use the space that was occupied by $S,T$ states for the output qubits.
The case of $k=3$ is displayed.
The big patch with an arrow illustrates that the size of that patch changes dynamically.
}
\includegraphics[width=0.48\textwidth, trim={0ex 45ex 115ex 0ex}, clip]{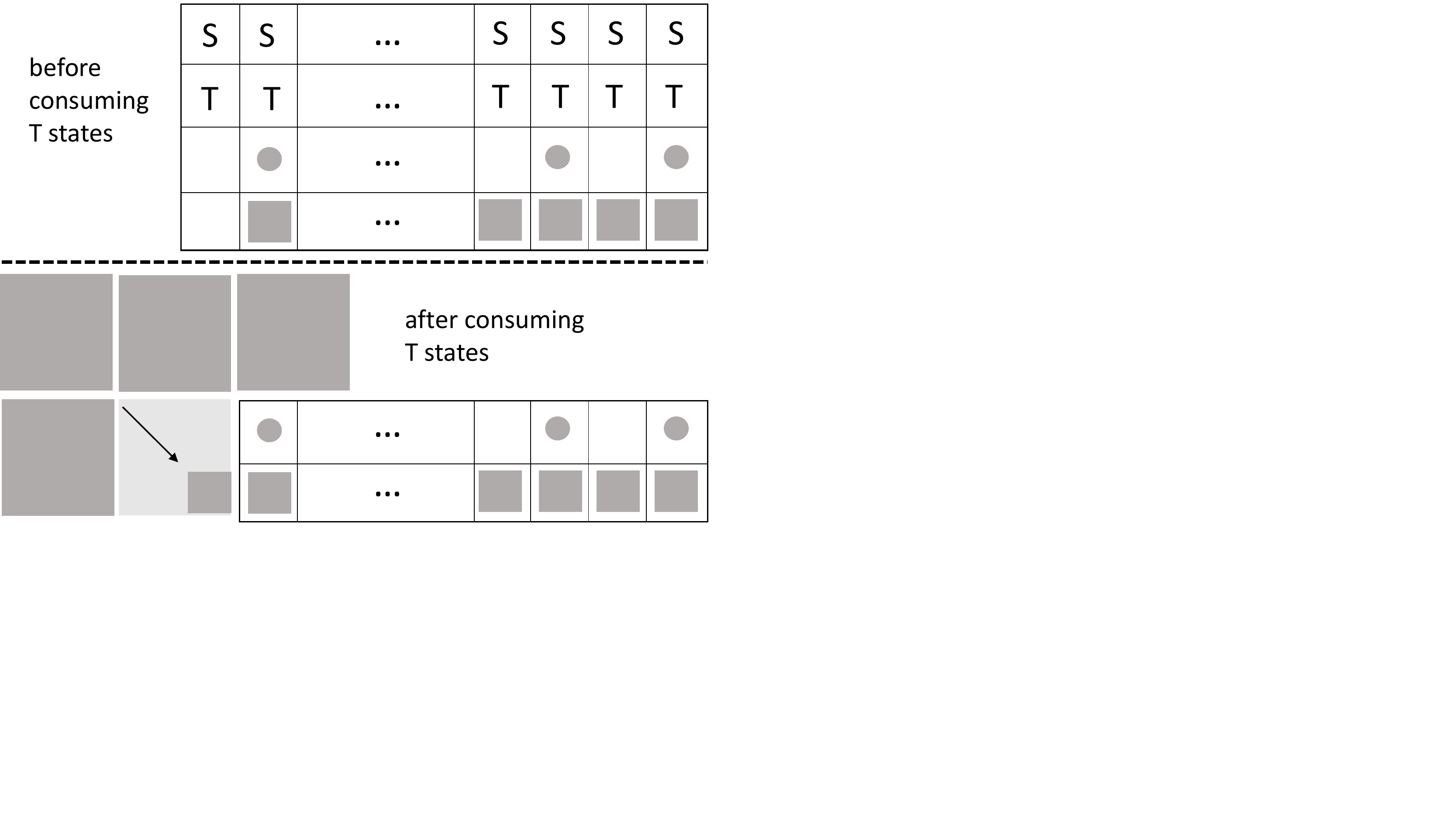}
\label{fig:layout}
\end{figure}

The overall layout of our protocol is depicted in \cref{fig:layout}.
For an $[[n,k,d]]$ block code,
there are $8n$ qubits that form a $4$-by-$2n$ rectangle.
The first and second rows are occupied by $T$ and $S$ states that are to be consumed.
Every other qubit of the third row is a data qubit of the block code.
The fourth row is reserved for cat state preparation.
If $T$ rotations are available in place, then of course we do not need the upper two rows.

Near the end of the array of qubits there will be $k+2$ qubits of better quality,
of which we have $k$ ``output qubits'' and two ancillas.
If we use surface code patches for qubits,
this means that the output qubits have larger size.
The precise size of these patches should be determined case-by-case.
The output qubits will interact with the encoded qubits of the block code 
only after all $S$ and $T$ states are consumed,
and will host distilled magic states.
Therefore, we may use the space occupied by the first and second row of the rectangle
for the output qubits.
Also, we can use the space of the leftmost column of the $4$-by-$2n$ rectangle
for the better quality qubits.

For quadratic error reductions, the size of the surface code patches for better quality qubits
would be roughly twice as big as the other qubits.
Precise sizes should be determined by desired quality of the output qubits.

\subsection{Error model}

Every measurement outcome is flipped with probability $p \in [0,1)$.
Every qubit suffers from independent noise after any operation
including the identity.
The single qubit error is modeled by a quantum channel 
\begin{align}
\mathcal E (\rho) = (1-p) \rho + p \mathcal D(\rho) \label{eq:qubitnoise}
\end{align}
where $\mathcal D$ is another quantum channel.
Note that we use the same $p$ for both the measurement outcome error 
and the single qubit noise.
Any $T$ states and any $T$ or $T^\dagger$ gates are
immediately followed by 
\begin{align}
\mathcal E_T (\rho) = (1-p_t) \rho + p_t \mathcal D(\rho), \label{eq:Tnoise}
\end{align}
but not by $\mathcal E$.
Note that we use the same $\mathcal D$ in \cref{eq:qubitnoise,eq:Tnoise}
for simplicity.

We will treat output qubits as if they were noise free;
using a surface code family
it suffices to choose a code distance that matches
the quality of the distilled magic states.

One may wonder how the independent noise assumption can be fulfilled
after measurements of $X^{\otimes m}$, a multiqubit operator.
A naive preparation of the cat states introduces correlated errors
and using such a cat state for our measurement will invalidate the
independent noise assumption.
Hence, we need a more careful protocol to prepare 
cat states that differ from ideal ones by independent noise
on constituent qubits.
The protocol in \cite{nicolas} achieves this goal.

\section{Protocols}

In the protocol specification below,
we put tildes on states and operators 
to emphasize that they are noisy
and are not as good as the output qubits.

\subsection{State teleportation to a large surface code patch}

In our protocol,
it is necessary to teleport a magic state that 
is encoded in a block code to an output qubit.
This will be performed by preparing a state of $Z_{out} = +1$
and measuring $X_{out} \mathbf X$ and $\mathbf Z$ 
where $\mathbf X, \mathbf Z$ are logical operators of the block code.
Using surface code patches,
it is straightforward to initialize the output qubit 
and measure $\mathbf Z$ of the code,
but the measurement of $X_{out} \mathbf X$ is unusual
since the data qubits of the block code have smaller size.
This problem is solved by a cat state preparation as the following.
Let $a, b$ denote two big patches in the bottom row of \cref{fig:layout}. Patch $b$ is the one with an arrow.
\begin{enumerate}
\item Prepare $\ket{++}_{ab}$ and measure $Z_a Z_b$ with a Pauli correction to have $\ket{00}_{ab} + \ket{11}_{ab}$.
\item Shrink $b$ to match the size of the patches in the fourth row of the $4$-by-$(2n-1)$ rectangle.
\item Following the prescription of~\cite{nicolas}, 
sufficiently measure $ZZ$ between nearest neighbors in the fourth row and discard every other qubit.
\end{enumerate}
According to \cite{nicolas},
fault-tolerant cat state preparation on a one-dimensional array of qubits
needs only one $ZZ$ measurement for the leftmost pair.
Our procedure exploits this construction,
and since the leftmost pair measurement is performed on big patches,
the output qubits will be protected as desired.
Once we have a cat state, we follow the procedure in \cref{eq:XmByCat}
to complete the $X_{out} \mathbf X$ measurement.

In all protocols below, only
one output qubit interacts with the block code at a time.
So, when we have multiple output qubits, 
we shift (teleport) the output qubit to the right after interacting.

\subsection{Second order $T$ distillation by $[[6,2,2]]$}

The stabilizers of the code $[[6,2,2]]$ that appear in~\cite{Jones2012multilevel}
are $XXXXII$, $IIXXXX$, $ZZZZII$, and $IIZZZZ$.
We choose logical qubits by two pairs of logical operators
$\mathbf X_1 = XIXIXI$, $\mathbf Z_1 = ZIZIZI$, $\mathbf X_2 = IXIXIX$, and $\mathbf Z_2 = IZIZIZ$.

\begin{enumerate}

\item Initialize six data qubits in $\ket{\tilde 0} ^{\otimes 6}$.
Bring $\ket{\tilde T}_{t1} \ket{\tilde T}_{t2}$ 
on the left of the array of data qubits,
where output qubits are yet to be set up.

\item Measure $XXXXII$ and $IIXXXX$ on the six data qubits.
Apply $Z$s such that the resulting state is the logical state with 
$\mathbf Z_1 \approx \mathbf Z_2 \approx +1$.

\item 
Teleport states in qubits $t1$, $t2$ into the logical qubits
by measuring $X_{t1} \mathbf X_1$, $X_{t2} \mathbf X_2$ 
and then measuring $Z_{t1}$, $Z_{t2}$,
followed by appropriate Pauli $Z$ corrections on the logical qubits.

\item Apply $\tilde T^{\otimes 6}$  on the data qubits.

\item Measure $X^{\otimes 6}$ \emph{twice} on the data qubits.
Postselect on all $+1$ outcomes.

\item Apply $(\tilde T^\dagger)^{\otimes 6}$ on the data qubits.

\item Initialize two output qubits $o1,o2$ in $\ket 0 ^{\otimes 2}$.
Measure $X_{o1}\mathbf X_1$ twice with outcomes $x_1, x_1'$,
and $X_{o2}\mathbf X_2$ twice with outcomes $x_2, x_2'$.
Postselect on consistent outcomes $x_1 = x_1'$ and $x_2 = x_2'$.
Apply $Z_{o1}$ if $x_1 = -1$ and $Z_{o2}$ if $x_2 = -1$.

\item Measure $XXXXII$ and $IIXXXX$ on the six data qubits. 
Postselect on all $+1$ outcomes.

\item Measure individual data qubits destructively in the $Z$ basis 
to obtain outcomes $z_1,\ldots,z_6$.
Postselect on both conditions $z_1 z_2 z_3 z_4 = +1$ and $z_3 z_4 z_5 z_6 = +1$.
Apply $X_{o1}$ if $z_1 z_3 z_5 = -1$ and $X_{o2}$ if $z_2 z_4 z_6 = -1$.

\item Accept the output qubits if all the postselections have succeeded.
The output qubits are in the distilled state $\ket T^{\otimes 2}$.

\end{enumerate}

Steps~1,2,3 are to prepare encoded $T$ states 
$\mathbf T_1 \mathbf T_2 \ket{\textbf{++}}$.
They have error rate $O(p)$.
Steps~4,5,6 are to measure the product of stabilizers 
$\mathbf M = 
(\mathbf T_1 \mathbf X_1 \mathbf T_1^\dagger)(\mathbf T_2 \mathbf X_2 \mathbf T_2^\dagger)$
of the encoded $T$ states.
The Clifford stabilizer $\mathbf M$ is induced by the transversal operator 
$(T^\dagger X T)^{\otimes 6} \propto S^{\dagger \otimes 6} X^{\otimes 6}$
because,
since $\mathbf X_{1,2}$ have weight $-1 \bmod 4$,
$S^{\dagger \otimes 6}$ induces $\mathbf S_1 \mathbf S_2$ on the code space.
Step~7 implements the half of a teleportation protocol.
Step~8 checks the $X$-stabilizers of the code.
Finally, Step~9 measures the $Z$-stabilizer as well as the $Z$-logical operator
to complete the teleportation of two logical qubits.

The measurement depth is counted as follows.
There are $m_1 = 3$ rounds of single-qubit measurements on the data qubits in Steps~1,3,9.
There are $m_2 = 2 + 2 + 2 + 2 = 8$ rounds of multiqubit $X$-measurements on the data qubits in Steps~2,3,5,8.
There are $2$ rounds of $T$-gates on data qubits in Steps~4,6, 
which would involve input $T$ and $S$ states.
An injection of $T$ gate that requires an $S$ correction involves $6$ one- and two-qubit measurements
(\cref{fig:injection}(c));
$T$ states can be transported right next to the data qubits and $S$ state right above $T$ state
while other steps are being executed.
So, there are $m_t = 12$ rounds of single- or two-qubit measurements.
There are $m_{out} = 4$ rounds of joint measurements in Step~3 between the output qubits and data qubits.
We may neglect the initialization of the output qubits
since that can be done in parallel with previous data qubits measurements.
Overall, we have 
$m_1 + m_t = 15$ rounds of one- and two-qubit measurements,
$m_2 = 8$ rounds of multiqubit $X$-measurements on the data qubits,
and $m_{out} = 4$ rounds $X$-measurements that involves the output qubits.

Let us be more specific on the measurement count using surface code patches,
treating one syndrome measurement of the surface code patch as the unit \emph{time}.
Assume that $d$ rounds of syndrome measurements are needed for surface code patches of the data qubits
and $d'$ rounds of syndrome measurements for the output qubits
for one logical operation; $d$ or $d'$ is the code distance of a patch.
To prepare a cat state that is ``2-fault-tolerant''~\cite{nicolas},
we need $8$ rounds of logical operations across nearest patches.
This time is long enough that we can ignore all explicit one- and two-qubit measurements in 
Steps~1,4,6 as they can run in parallel with
the cat state preparation for the next step. 
Step~9 is also negligible.
The last round of the cat state preparation of \cite{nicolas}
consists of single-qubit $X$ measurements on qubits that do not partake in the final cat state,
which can run in parallel with $XX$ measurements between the cat state qubits and data qubits.
Including single-qubit $Z$ measurements in \cref{eq:XmByCat},
we conclude that one multiqubit $X$-measurement takes time $9d$,
or $7d + 2d'$ if it involves the output qubits.
So, the total time is $m_2 \cdot 9d + m_{out}(7d + 2d')$.

The number of physical qubits used is $4 \cdot 12 \cdot d^2 + 4 d'^2$,
neglecting ancillas for syndrome measurements of the surface code.

\subsection{Third order $T$ distillation by $[[7,1,3]]$}

The following is our measurement sequence 
based on the principle of measuring $TXT^\dagger$
that has eigenvalue $+1$ on $\ket T$~\cite{Knill2004a,Knill2004b,HHPW}.
An implementation circuit of this idea was also presented in \cite{AliferisGottesmanPreskill2005}
but without consideration of a separate output qubit and limited connectivity.
The stabilizers of the Steane code $[[7,1,3]]$ are
$XIXIXIX$, $IXXIIXX$, $IIIXXXX$, $ZIZIZIZ$, $IZZIIZZ$, and $IIIZZZZ$.
The logical operators are $\mathbf X = X^{\otimes 7}$ and $\mathbf Z = Z^{\otimes 7}$.

\begin{enumerate}
\item Initialize seven data qubits in $\ket{\tilde 0} ^{\otimes 7}$.
Bring $\ket{\tilde T}_t$ on the left of the data qubit row,
where an output qubit is yet to be set up.

\item Measure the three $X$-stabilizers $XIXIXIX$, $IXXIIXX$, $IIIXXXX$ on the seven data qubits.
Apply $Z$s such that the resulting state is the logical state with $\mathbf Z \approx +1$.

\item Teleport $\ket{\tilde T}_t$ into the code by measuring $X_t \mathbf X$ and then $Z_t$.

\item Measure the three $X$-stabilizers. Postselect on all $+1$ outcomes.

\item Apply $\tilde T^{\otimes 7}$ on the data qubits.

\item Measure $X^{\otimes 7}$ twice. Postselect on all $+1$ outcomes.

\item Apply $\tilde T^{\dagger \otimes 7}$ on the data qubits.

\item Initialize an output qubit in the state $\ket 0_{o}$.

\item Measure each of three equivalent $X$-logical operators of the code,
multiplied by the $X$-operator on the output qubit:
$X_o(IXIXIXI)$, $X_o(XIIXXII)$, $X_o(XXXIIII)$.
Let $x_{1,2,3} = \pm 1$ be the outcomes.
Postselect on consistent results $x = x_1 = x_2 = x_3$.
Apply $Z_o$ if $x = -1$.

\item Measure the three $X$-stabilizers. Postselect on all $+1$ outcomes.

\item Destructively measure all data qubits in the $Z$ basis with outcomes $z_1,\ldots,z_7$.
Postselect on three conditions $z_1 z_3 z_5 z_7 = +1$, $z_2 z_3 z_6 z_7 = +1$,
and $z_4 z_5 z_6 z_7 = +1$.
Apply $X_o$ if $z_1 z_2 z_3 = -1$.

\item Accept the output qubits if all the postselections have succeeded.
The output qubit holds distilled $T$ state.
\end{enumerate}

Steps~1,2,3 prepare an encoded $T$ state to error rate $O(p)$.
Step~4 checks $X$-stabilizers.
Steps~5,6,7 measure the Clifford stabilizer $\mathbf T \mathbf X \mathbf T^\dagger$
of the encoded $T$ state.
Similar to the previous protocol,
the Clifford stabilizer is induced by $(T^\dagger X T)^{\otimes 7}$
since the logical operator $\mathbf X$ has weight $-1 \mod 8$.
Note that Step~4 has no analog in the previous protocol with the quadratic error reduction.
Step~4 here is needed because, without it, a two-error process,
where one $Z$ error in Steps~1,2,3 and another in Step~7,
would cancel each other to let an incorrect $T$ state pass
through the Clifford stabilizer check.
Steps~8,9,10,11 combine teleportation of the encoded $T$ state with 
Pauli stabilizer checks.
An interesting point in Step~9 is that it uses three representatives of logical operators
to detect second order processes that may result in wrong teleportation.

The measurement depth is counted as follows.
As in the time analysis of the previous protocol,
we just count the number of multiqubit measurements.
We need ``$3$-fault-tolerant'' cat states,
for the preparation of which we need $10$ rounds of one- and two-qubit measurements.
For those that do not involve the output qubits,
there are $3 + 3 + 2 + 3 = 11$ measurements in Steps~2,4,6,10.
For those that involve the output qubits,
there are $1 + 3 = 4$ measurements in Steps~3,9.
Using surface code patches of distance $d$ (data) and $d'$ (output),
we have $11\cdot 11d + 4 \cdot (9d + 2d')$ rounds of syndrome measurements
for the surface code patches.

The number of qubits used is $56d^2 + 3d'^2 $, neglecting ancillas of the surface code.

\subsection{Second order $CCZ$ distillation by $[[8,3,2]]$}

The following protocol is based on a generalization~\cite{campbell2017unified}
of triorthogonal codes~\cite{bh} that induces a logical CCZ gate
upon transversal $T$ gate.
We use a code with stabilizers $XXXXXXXX$, $IZIZIZIZ$, $IIZZIIZZ$,
and $IIIIZZZZ$.
We choose logical operators as
$\mathbf X_1 = IXIXIXIX$, $\mathbf Z_1 = IIIIIIZZ$, 
$\mathbf X_2 = IIXXIIXX$, $\mathbf Z_2 = IIIIIZIZ$,
$\mathbf X_3 = IIIIXXXX$, and $\mathbf Z_3 = IIIZIIIZ$.

\begin{enumerate}
\item Initialize eight data qubits in $\ket{\tilde 0}^{\otimes 8}$.

\item Measure the $X$-stabilizer $XXXXXXXX$ 
as well as the three $X$-logical operators $\mathbf X_{1,2,3}$.
Upon $-1$ outcomes, apply Pauli corrections by $ZIIIIIII$
and $Z$-logical operators $\mathbf Z_{1,2,3}$
such that the resulting state is the logical state $\ket{\tilde + \tilde + \tilde +}$.

\item Measure $IXXIXIIX$, the product of the three $X$-logical operators.
Postselect on the $+1$ outcome.

\item Apply $\tilde T^\dagger \otimes \tilde T \otimes \tilde T \otimes \tilde T^\dagger
\otimes \tilde T \otimes \tilde T^\dagger \otimes \tilde T^\dagger \otimes \tilde T$
on the data qubits. (We inserted $\otimes$ for clearer reading.)

\item Initialize three output qubits in the state $\ket{000}_{o1,o2,o3}$.
Measure each of $X_{o1} \mathbf X_1$, $X_{o2} \mathbf X_2$, and $X_{o3} \mathbf X_3$ twice,
with outcomes $x_1,x_1',x_2,x_2',x_3,x_3'$.
Postselect on consistent outcomes $x_j = x_j'$ for $j=1,2,3$.
Apply $Z_{oj}$ if $x_j = -1$ for $j = 1,2,3$.

\item Measure the $X$-stabilizer $X^{\otimes 8}$ on the data qubits. 
Postselect on the $+1$ outcome.

\item Destructively measure all data qubits in $Z$ basis with outcomes $z_1,\ldots,z_8$.
Postselect on all four conditions 
$z_1 z_2 z_3 z_4 = +1$,
$z_2 z_4 z_6 z_8 = +1$, 
$z_3 z_4 z_7 z_8 = +1$, and
$z_5 z_6 z_7 z_8 = +1$.
Apply $X_{o1}$ if $z_7 z_8 = -1$, $X_{o2}$ if $z_6 z_8 = -1$,
and $X_{o3}$ if $z_4 z_8 = -1$.

\item Accept the output qubits if all the postselections have succeeded.
The output qubits are in a distilled state~$\ket{CCZ}$.
\end{enumerate}

In Step~4, we choose a particular product of $T$ and $T^\dagger$.
This choice ensures that the underlying (generalized) triorthogonal code
satisfies ``level-3'' orthogonality~\cite{Haah2017tower},
which removes the need of Clifford corrections of \cite{campbell2017unified}.

The measurement depth is counted similarly as the previous protocols;
we just count multiqubit measurements.
We need ``2-fault-tolerant'' cat states.
There are $4 + 1 + 1 = 6$ measurements in Steps~2,3,6 that do not involve the output qubits.
There are $6$ measurements in Step 5 that involve output qubits.
Using surface code patches of distance $d$ (data) and $d'$ (output),
the duration of the protocol is $6\cdot 9d + 6 \cdot (7d + 2d')$.

The number of qubits used is $4 \cdot 16d^2 + 5 d'^2$,
neglecting ancillas of the surface code.

\section{Fault-tolerance analysis}

\begin{table}[h]
\centering

\caption{
Error of the output states measured in the trace distance
$p_{out} = \frac 1 2 \left\| \rho_{ideal} - \rho_{out} \right\|_1$
as a function of two parameters $p, p_t$ that represent 
the strength of noise for Clifford operations and $T$ gate/states, respectively.
}

\begin{tabular}{c|c|c|c}
\hline \hline
\multicolumn{4}{c}{$T$-to-$T$ using $[[6,2^\star,2]]$}\\
\multicolumn{4}{c}{$p_{accept} \ge (1-p)^{91} (1-p_t)^{14}$}\\
\hline
$\mathcal D(\rho)$ & $p_t^2$ & $p_t p$ & $p^2$ \\
\hline
$\frac 1 3(X\rho X + Y \rho Y + Z \rho Z)$ & 2 & 32 & 84 \\
$X \rho X$ & 6 & 62 & 188 \\
$Z \rho Z$ & 7 & 89 & 97 \\
\hline
\end{tabular}
\\
$^\star$ We do not take reduced density matrices on each output qubit when computing $p_{out}$.

\vspace{5ex}

\begin{tabular}{c|c|c|c|c}
\hline\hline
\multicolumn{5}{c}{$T$-to-$T$ using $[[7,1,3]]$}\\
\multicolumn{4}{c}{$p_{accept} \ge (1 - p)^{141}(1-p_t)^{15}$}\\
\hline
$\mathcal D(\rho)$ & $p_t^3$ & $p_t^2 p$ & $p_t p^2$ & $p^3$ \\
\hline
$\frac 1 3(X\rho X + Y \rho Y + Z \rho Z)$ & 3 & 34 & 276 & 855 \\
$X \rho X$ & 16 & 335 & 2606 & 6977\\
$Z \rho Z$ & 35 & 183 & 429 & 355 \\
\hline
\end{tabular}

\vspace{5ex}

\begin{tabular}{c|c|c|c}
\hline \hline
\multicolumn{4}{c}{$T$-to-$CCZ$ using $[[8,3,2]]$}\\
\multicolumn{4}{c}{by Step~7a; see the text.}\\
\multicolumn{4}{c}{$p_{accept} \ge (1-p)^{124} (1-p_t)^{8}$}\\
\hline
$\mathcal D(\rho)$ & $p_t^2$ & $p_t p$ & $p^2$ \\
\hline
$\frac 1 3(X\rho X + Y \rho Y + Z \rho Z)$ & 8 & 171 & 958\\
$X \rho X$  & 18 & 397 & 2340 \\
$Z \rho Z$ & 28 & 462 & 1414 \\
\hline
\end{tabular}

\label{tb:pout}
\end{table}

We simulated the complete protocols using density matrices
which is easy as they involve at most 11 qubits.
We numerically examined $p_{out}$ as a function of $p,p_t \in (10^{-6}, 10^{-4})$
for a given $\mathcal D$,
and fitted to a polynomial formula 
$p_{out} = a p_t^2 + b p_t p + c p^2$
or $p_{out} = a p_t^3 + b p_t^2 p + c p_t p^2 + d p^3$
where $a,b,c,d$ are fitting parameters.
\Cref{tb:pout} shows these coefficients rounded to integers.

The $CCZ$ state distillation protocol is designed to achieve quadratic error suppression,
and thus all the cat states there need to be $2$-fault-tolerant only~\cite{nicolas}.
Assuming that indeed the cat states are only $2$-fault-tolerant,
it is not too meaningful to analyze our protocol 
where Step~7 is supposed to achieve quartic error suppression for $X$ errors.
Hence, to have a conservative estimate of $p_{out}$ in the $CCZ$ distillation protocol,
we use the following alternative to Step~7 in our simulation:
\begin{enumerate}
\item[7a.]
Destructively measure all data qubits in $Z$ basis with outcomes $z_1,\ldots,z_8$.
Postselect on the condition $z_1 z_2 z_3 z_4 z_5 z_6 z_7 z_8 = +1$.
Apply $X_{o1}$ if $z_7 z_8 = -1$, $X_{o2}$ if $z_6 z_8 = -1$,
and $X_{o3}$ if $z_4 z_8 = -1$.
\end{enumerate}
In practice, Step~7 should be preferred to this alternative;
Step~7 will catch more errors with no complication of quantum operations in comparison to~7a.
The overall success probability decreases when we do Step~7 rather than 7a
but only by a negligible amount.
If one is curious what $p_{out}$ would be using Step~7, not~7a,
under the very noise model of ours, we provide the following formulas:
$p_{out} \approx 3 p_t^2 + 58 p_t p + 200 p^2$ if $\mathcal D(\rho) = \frac 1 3 (X\rho X + Y \rho Y + Z \rho Z)$,
$p_{out} \approx 28 p_t^2 + 462 p_t p + 1397 p^2$ if $\mathcal D(\rho) = Z \rho Z$,
and
$p_{out} \approx 12 p_t^4 + 0.035 p_t p + 9 p^2 + 2300 p^3$ if $\mathcal D(\rho) = X \rho X$.

To test the accuracy of the leading order formulas
for $p_{out}$ that are presented in \cref{tb:pout},
i.e., the contributions from higher order terms,
we computed $p_{out}$ as a function of $\lambda = 0.0, 0.2, 0.4, 0.6, 0.8, 1.0$
by setting $p = 10^{-2}\lambda$ and $p_t = 10^{-2} (1-\lambda)$.
We observed that the leading order formulas for $p_{out}$
are correct to $29\%$ relative accuracy for these values of $p,p_t$ in all cases.


Note that conventional leading order formulas are reproduced
in our analysis by setting $p=0$ and $\mathcal D(\rho) = Z\rho Z$:
(i) $p_{out} \approx 7p_t^2$ for $[[6,2,2]]$~\cite{jones2013low} 
which is equivalent~\cite{Haah2017tower} to the smallest example of 
triorthogonal codes~\cite{bh},
(ii) $p_{out} \approx 35 p_t^3$ for $[[7,1,3]]$~\cite{Knill2004b}
which is equivalent~\cite{Haah2017tower} 
to the smallest example of quantum Reed-Muller 
codes~\cite{BravyiKitaev2005Magic,rtrio},
and
(iii) $p_{out} \approx 28 p_t^2$ for $[[8,3,2]]$~\cite{eastin2013distilling,jones2013low,campbell2017unified,rtrio}.

The acceptance probability $p_{accept}$ was also numerically computed,
and the formula was satisfied by all the three error channels considered.
The exponents of $(1-p)$ and $(1-p_t)$ are approximately the number of possible error locations
under our error model.
Note that this acceptance probability assumes that the cat state preparation is successful.

\section{Surface Code: Error Detection or Partial Error Correction}
Typically, the surface code in a patch is assumed to operate in an ``error correcting'' mode, meaning that one attempts to correct any errors that occur.  At the same time, typically one assumes that the outermost code is used in an ``error detecting'' mode, meaning that one only keeps the distilled magic state if {\it no} errors are detected.  
A final possibility is ``partial error correction''; 
for example, in \cite{rtrio}  it was suggested that for certain large outer codes one
might correct if  a {\it small} number of errors would give the observed syndrome, 
and discard otherwise. 
More generally, one may choose some set of observed syndromes to correct and discard on others.

We now consider error detection and partial error correction as applied to surface codes
inside a magic state factory (where if one has to discard the state, then this has no effect on the rest of the computation).
It is an interesting question whether partial error correction might be useful on the qubits used inside a quantum computer {\it outside the magic state factory}, i.e., on the qubits actually used for computation.  In this case, if one has to discard the state on some observed syndrome, this will typically require discarding all of the computation up to that point, since typically that qubit will be entangled with the rest of the computation, and restarting the computation from scratch.  Thus, for this to be useful, the probability of discarding the state would have to be small compared to the inverse number of gates in the computation.

In error detecting mode, meaning that we discard the state whenever any error in the surface code is detected, a surface code of distance $d$ can now suppress errors up to $d$-th order:
by performing $d$ rounds of syndrome measurements after each logical measurement, we can suppress logical errors
if fewer than $d$ physical errors occur in any round of logical measurements.
Unfortunately, this simple error detection mode may not be too useful.
The average number of errors is $p$ times the number of error locations; there are $d^2-1$ syndromes, and so we perform $d(d^2-1)$ syndrome measurements.  Each syndrome measurement is broken into some number of physical operations, with the exact number depending in detail on the physical implementation.  So, one needs $p d(d^2-1) \ll 1$ to attain a large probability that the state will not be discarded on a given round, with the exact value depending on the implementation of syndrome measurements.  

For a $[[7,1,3]]$ code, we need $\sim 60$ patches.
If $d=3$ for the surface code,
given a total number of rounds 
$\sim 100$, we need $6000 \cdot d(d^2-1)\approx 1.6 \times 10^{5} \ll p^{-1}$
to obtain significant throughput
(there are some additional numerical factors of order $1$ due to use of ancillas to implement measurements).  
So for physical error rate $p \approx 10^{-5}$,
it is unlikely that no error is detected.
The chance of acceptance will be worse if the distance of the surface code is larger.
On the other hand, 
if $p$ is so small (say $p < 10^{-6}$) 
that we have high success probability with code patches in the detecting mode,
then for $p_t \approx 10^{-2}$
there is no need to use the surface code patch 
because then the dominant error term is the term of order $p_t^3$, 
rather than higher order terms~$p_t^2 p, p_t p^2, p^3$ in~$p$.

Partial error correction, in which one corrects for example up to one error {\it in each patch, in each round}, may be more likely to succeed.  In contrast to error detection, we only discard a patch if two errors occurs, which occurs with probability
roughly ${d(d^2-1) \choose 2} p^2 \approx (d^3 p)^2/2$.  Given $\sim 60$ patches and $\sim 100$ rounds, we need
$6000 \cdot (d^3 p)^2 \ll 1$, or equivalently
$\sqrt{6000} d^3 p\ll 1$, which is much more attainable.  However, analyzing the performance of partial error correction will require an enumeration of error patterns that we leave for future work.  At even higher physical noise rates, it may be useful to implement even more relaxed forms of partial error correction in which one corrects larger numbers of errors.  We also leave this for future work.

\bibliographystyle{apsrev4-1}
\nocite{apsrev41Control}
\bibliography{small-ref}
\end{document}